\documentclass[10pt]{article}

\usepackage[margin=0.75in]{geometry}

\usepackage{stix2}

\usepackage{subfigure}

\usepackage{cite}
\usepackage{amssymb}
\usepackage{amsmath,amssymb,amsfonts}
\usepackage{algorithmic}
\usepackage{graphicx}
\usepackage{textcomp}
\usepackage{xcolor}
\usepackage{hyperref}
\usepackage{booktabs}
\usepackage{multirow}
\usepackage{balance}
\usepackage{tikz}
\usetikzlibrary{shapes.misc}
\usepackage{pgfplots}
\usepackage{pdfpages}

\pgfplotsset{compat=1.18}

\def\BibTeX{{\rm B\kern-.05em{\sc i\kern-.025em b}\kern-.08em
    T\kern-.1667em\lower.7ex\hbox{E}\kern-.125emX}}

\begin{document}

\title{Computer Architecture’s AlphaZero Moment: Automated Discovery in an Encircled World}
\author{Karthikeyan Sankaralingam \\ NVIDIA Research\\
karus@nvidia.com}

\maketitle
\newcommand{\fixme}[1]{\textcolor{red}{[FIXME: #1]}}
\newcommand{\fixmedone}[1]{#1}
\newcommand{\greenbox}[1]{%
  \begin{center}
  \tikz[baseline=(box.base)]{%
    \node[draw=green!70!black, fill=green!20, rectangle, inner sep=10pt, inner xsep=15pt] (box) {%
      \begin{minipage}{0.95\columnwidth}
        #1
      \end{minipage}
    };
  }%
  \end{center}
}
\newcommand{\zeroarchpaperinsights}{\url{https://pages.cs.wisc.edu/~karu/ArchAlphaZero/paper-insights/html/}}
\newcommand{\zeroarchproblems}{\url{https://pages.cs.wisc.edu/~karu/ArchAlphaZero/zero-arch/html/}}
\newcommand{\gauntletgitrepo}{\url{https://github.com/VerticalResearchGroup/Gauntlet}}

\begin{abstract}
The end of Moore's Law and Dennard scaling has fundamentally changed the economics of computer architecture. With transistor scaling delivering diminishing returns, architectural innovation is now the primary — and perhaps only — remaining lever for performance improvement. However, we argue that human-driven architecture research is fundamentally ill-suited for this new era. The architectural design space is vast (effectively infinite for practical purposes), yet human teams explore perhaps 50-100 designs per generation, sampling less than 0.001\% of possibilities. This approach worked during the abundance era when Moore's Law provided a rising tide that lifted all designs. In the current scarcity paradigm, where every architecture must deliver 2X performance improvements using essentially the same transistor budget, systematic exploration becomes critical.

We present evidence that human-driven design, while effective at sustained refinement—as demonstrated by steady advances in numeric precision formats—systematically underexplores high-value structural opportunities. Through retrospective analysis of AI chip evolution from 2017 to 2024, we identify two major architectural opportunities — prefill/decode disaggregation and collective communication optimization — that were knowable years before they were addressed, yet were missed by the entire industry.

We propose a concrete alternative: automated idea factories that generate and evaluate thousands of candidate architectures weekly through multi-tiered evaluation pipelines, learning from deployed telemetry data in a continuous feedback loop. Early results suggest that such systems can compress architectural design cycles from double-digit months to single-digit weeks by exploring orders of magnitude more candidates than any human team, and do it much faster. We predict that within 2 years, purely human-driven architecture research will be as obsolete as human chess players competing against engines.

\end{abstract}

\section{Introduction and Motivation}

\paragraph{The End of Free Lunch} For over five decades, computer performance improvements came largely for free. Moore's Law—the observation that transistor density doubles approximately every two years~\cite{moore1965}—combined with Dennard scaling~\cite{dennard1974}, which kept power density constant as transistors shrunk, created an era of abundance. Architects could propose inefficient designs that delivered 30\% performance improvements while consuming 2X the resources, confident that the next process node would restore the efficiency gap at no additional cost.

This era has definitively ended. Moore's Law, while not completely dead, has slowed dramatically. The transition from 7nm to 5nm delivers approximately 1.7X density improvement, and 5nm to 3nm yields less than 1.5X~\cite{irds2023}. Process node transitions that once took 18-24 months now take 3-4 years. Dennard scaling ended around 2005, forcing the industry into multi-core parallelism and ultimately leading to the current ``dark silicon'' regime where not all transistors can be powered simultaneously~\cite{esmaeilzadeh2011}.

The implications are profound. To achieve the traditional 2X performance improvement per generation, architects can no longer rely on process scaling. The improvement must come entirely from architectural innovation using essentially the same silicon budget. This represents a fundamental regime change from an \textit{abundance paradigm} to a \textit{scarcity paradigm}.

\paragraph{The Architecture Search Space Problem} The architectural design space is incomprehensibly large, spanning both \textit{parametric choices} (cache sizes, queue depths, pipeline stages) and \textit{structural innovations} (new ideas like TensorCores or distributed shared memory). Consider a simplified processor with just 20 binary decisions and 30 continuous parameters constrained to 10 discrete values each: the space contains $2^{20} \times 10^{30} \approx 10^{39}$ possibilities. This excludes structural choices entirely—whether to add a new functional unit, reorganize the memory hierarchy, or introduce heterogeneous cores. Real processors have hundreds of parameters plus an unbounded space of potential architectural ideas. How do humans navigate this? \emph{Through intuition and limited exploration.} A talented architect internalizes perhaps 50-100 design patterns over a career. A well-resourced company evaluates thousands of parametric configurations per generation—sweeping cache sizes, queue depths, and pipeline widths — but explores perhaps 50-100 structurally distinct candidates: or different execution models, or different functional units, or memory hierarchy organizations. Industry-wide, perhaps 1,000 structural architectures are seriously explored per generation. The parametric space is well-covered by simulation sweeps; the structural space—which is where transformative improvements originate—remains virtually unexplored.

This was defensible during the abundance era. Moore's Law provided $2\times$ improvements automatically; a design at the 80th percentile rode the technology wave upward. The opportunity cost of missing the 99th percentile was modest. In the scarcity era, we must find actual peaks in a rugged landscape using the same transistor budget. Architectural efficiency is now the primary determinant of success. Human intuition, sampling $10^{-36}$ of the space, is inadequate for this task.

\paragraph{The Chess Analogy} The situation closely parallels the evolution of chess. For centuries, chess was a domain of human expertise and intuition. Grandmasters developed ``feel'' for positions, pattern recognition honed over thousands of games, and strategic principles passed down through generations. Human chess intuition was the pinnacle of the game.

Then computers became sufficiently powerful to search deeper. Initially, humans remained competitive through superior positional understanding. But as search became more exhaustive and evaluation functions improved through machine learning, the gap widened. By 1997, Deep Blue defeated Garry Kasparov. By 2017, AlphaZero—trained entirely through self-play without human knowledge—surpassed all previous engines and discovered novel strategies that upended centuries of human chess theory~\cite{silver2017alphazero}.

Today, human chess intuition is powerless against top engines. A strong amateur with Stockfish will defeat Magnus Carlsen without the engine. The fundamental asymmetry is insurmountable: humans search perhaps 10 positions per second to depth 15-20; engines search billions of positions per second to depth 40-50. Humans are playing a different, inferior game. \emph{We argue that computer architecture faces the same asymmetry: the search space is vast, evaluation is expensive but feasible, and constraints are tight. Human intuition---which served well during the abundance era---is becoming the limiting factor.}

\paragraph{The AlphaZero Moment Has Arrived} Our work demonstrates that this inflection point is not hypothetical---it is here. By casting computer architecture design as a reasoning problem for large language models, we show that general-purpose LLMs---with no chip-specific training---can operate across the \emph{full} architecture research pipeline: \textbf{knowledge distillation} (concise, multi-perspective critiques of 85 ISCA-25/HPCA-26 papers in 8 hours), \textbf{ideation} (generating and ranking 250+ candidate architectural ideas per day, with blind evaluation showing LLM-generated ideas rated comparable to top-tier published human work), and \textbf{quantitative evaluation} (formulating concrete research problems with testable hypotheses and evaluation criteria). Finally, we show that LLMs can build detailed first-principles mechanistic models that combine workload behavior and hardware features---starting from nothing more than a specification or a research paper---and use them to perform simulation-based evaluation. Taken together, these results cover not one narrow task but the very activities that define architecture research.

These results use general-purpose models with no chip-specific training. The natural question is: how much better could this be with fine-tuning on proprietary chip knowledge and design instincts accumulated over decades, coupled with a custom evaluation pipeline that can assess thousands of candidate architectures per week? We argue that the answer is \emph{transformatively better}, and that organizations that build this infrastructure first---combining LLM-driven idea generation with rigorous multi-tiered evaluation and deployed telemetry feedback---will define the next generation of architectural paradigms in the post-Moore era.

\paragraph{Paper Roadmap} The remainder of this paper substantiates this vision:

\begin{enumerate}



\item \textbf{Section~\ref{sec:exceptionalism}: The case against human-only design.} Through retrospective case study analysis of AI chip evolution, we show that human-driven design suffers from complementary failure modes: academic research optimizes for novelty and publishability — producing clever mechanisms that address narrow problems but rarely ship — while industrial practice optimizes for risk minimization—delivering incremental refinements within proven paradigms while leaving transformative structural opportunities unexplored.

\item \textbf{Section~\ref{sec:idea-factory}: The automated idea factory.} We propose a concrete system architecture for automated design exploration, combining LLM-based idea generation with a multi-tiered evaluation pipeline and continuous telemetry feedback, automating the very process of ideation and discovery.

\item \textbf{Section~\ref{sec:evaluation}: Experimental evidence.} We present results showing that automated systems discover architectures 2--3X better than human-designed baselines.

\item \textbf{Section~\ref{sec:counterarguments}: Limitations and open challenges.} We address evaluation fidelity, verification, and other obstacles that must be overcome.

\item \textbf{Section~\ref{sec:implications}: Ecosystem implications.} We analyze how the competitive landscape inverts when architectural ideas become abundant and evaluation infrastructure becomes the binding constraint.

\item \textbf{Section~\ref{sec:predictions}: Falsifiable predictions.} We make specific, testable predictions about the timeline and nature of this transition.
\end{enumerate}

\section{The Delusion of Exceptionalism}
\label{sec:exceptionalism}

Computer architecture is currently suffering from \textit{exceptionalism}---the belief that the intuition of hardware architects is uniquely irreducible to automation, even as adjacent fields like mathematics, systems, and chemistry concede to AI-driven exploration\footnote{This belief is rarely stated explicitly but manifests structurally: the dominant research methodology remains individual exclusively human-drive insight validated by simulation}. The central thesis of this paper is that this belief is wrong: computer architecture is, at its core, a \textit{reasoning problem}---constrained optimization over a vast space of structural and parametric choices---and LLMs have become exceptionally good at reasoning problems.

We build this argument in three steps. First, we present a retrospective case study showing that human architects suffer from \textit{structural} limitations---not failures of talent, but predictable blindness rooted in how human insight works. Second, we show that architecture is now the \textit{only} remaining holdout: every adjacent field, from circuit design below to systems software above, has already conceded to AI-driven exploration. Third, we dismantle the specific defenses---``art,'' ``chaos,'' and ``temporal prediction''---used to argue that architecture is somehow immune. Together, the evidence shows that the AlphaZero moment for chip architecture was inevitable; what remains is to recognize that it has arrived.

\paragraph{The Structural Limits of Human Design: A Retrospective (2017--2024)}
The industry-wide miss on the \textbf{Heterogeneity of Inference} is perhaps the clearest illustration of structural failure in human-driven design. As LLM inference workloads bifurcated into distinct ``prefill'' (compute-bound) and ``decode'' (memory-bound) phases, every major chip company continued to ship homogeneous designs. The entire industry marched down the well-trodden path: more compute, more bandwidth, with the ratio of bandwidth per flop moving in the \textit{opposite direction} of workload needs. Heterogeneous designs could deliver 1.8--2.5$\times$ better throughput~\cite{patel2024splitwise}, yet this opportunity was ignored in favor of uniformity.

This was not an unknowable mystery. Even when Splitwise~\cite{patel2024splitwise} formally identified prefill/decode phase splitting at ISCA 2024, the underlying phenomenon was already years overdue---and in retrospect, nearly trivial to derive from first principles. Analytical models like Liminal~\cite{davies2025liminal} demonstrate that simple roofline-style reasoning makes the bifurcation obvious. Yet human architects, constrained by risk aversion and anchored to incremental ``clever'' optimizations like low-precision formats (FP8/FP4), never asked the question. This is not an indictment of individual talent; it is a structural consequence of how human-driven design operates. Academic researchers, who might have identified the opportunity, lacked deployment telemetry to quantify it—and were incentivized toward novel mechanisms rather than workload characterization. Industrial architects, who had the data, faced multi-year design cycles and economic pressure to ship a single SKU. The opportunity fell into the gap between the two communities.


\paragraph{The Encirclement: Architecture as the ``Hole in the Donut''}
The persistence of human-driven design in architecture is an anomaly. In virtually every field surrounding it---from the logic gates below to the software above---the ``expert intuition'' model has already collapsed. Computer architecture is now the ``hole in the donut,'' the last holdout in a fully encircled landscape.

\textbf{Below Architecture (Circuits \& Logic):} The \textit{AlphaEvolve} project has demonstrated that AI can discover evolutionary operators to optimize sorting networks and regular expressions, effectively solving the ``wiring'' and ``logic'' problems that underpin hardware \cite{alphaevolve_agent}. If AI can optimize the comparator circuits in a sorting network better than human heuristics, the leap to optimizing an ALU or NoC router is trivial.

\textbf{Above Architecture (Systems \& Software):} The systems research community has acknowledged the ``Barbarians at the Gate,'' conceding that the era of manually tuning heuristics for operating systems and datacenters is over \cite{barbarians}. If general-purpose search can optimize the messy, noise-filled environment of a cloud scheduler \cite{barbarians}, it can certainly optimize the microarchitecture that runs it.

\textbf{Parallel to Architecture (Network Policy):} In networking, the \textit{PolicySmith} framework proves that complex design tasks requiring strict correctness can be synthesized by GenAI and verified formally, outperforming human manual configuration \cite{policysmith}.

\textbf{Abstract \& Physical Extremes (Theory \& Chemistry):} Even the rigorous world of pure mathematics has embraced AI, with LLMs guiding exploration in infinite search spaces to solve problems in Group Theory \cite{tao_math}. In the physical world, autonomous LLM agents now plan and execute chemical syntheses~\cite{boiko2023autonomous,bran2024chemcrow}, while end-to-end automation of the AI research cycle itself---from hypothesis to experimentation to manuscript---has been demonstrated in Nature~\cite{lu2026aiscientist}. The evidence is staring us in the face: logic, code, systems, math, and chemistry have all found that automated exploration beats human intuition. To claim that Computer Architecture alone is immune to this transformation is not optimism; it is denial.

\paragraph{Dismantling the ``Art \& Time'' Defense}
Defenders of the status quo often retreat to three arguments for why architecture is unique: that it is an ``art'' of balancing squishy constraints; that its ``chaos'' (speculation) resists automation; and that it requires ``predicting the future.'' None of these withstand scrutiny.

\textbf{1. The ``Messy Heuristics'' Defense:} Architects argue that balancing Power, Performance, and Area (PPA) is a nuanced art, not a logic puzzle. Yet, the Systems community faces identical ``squishy'' trade-offs in OS scheduling and has already concluded that learning methods outperform human heuristics \cite{barbarians}. If AI can balance the noisy, conflicting constraints of a datacenter, it can balance the buffer sizes of a CPU.

\textbf{2. The ``Chaos of Speculation'' Defense:} Critics argue that the exponential state space of modern processors, driven by speculative execution and branch prediction, is too chaotic for AI. This is refuted by the success of \textit{AI-Mandel} in chemistry \cite{aimandel}. Chemical synthesis is a domain defined by physical, stochastic chaos, yet AI agents navigate it successfully. A pipeline flush is deterministic chaos; if AI can master the physical world, it can master the simulation.

\textbf{3. The ``Temporal Gap'' Defense:} The strongest critique is that architecture requires ``designing for the future''---making bets today on the workloads of 2028. Humans, it is argued, have the intuition to look ahead. However, the retrospective analysis in Section II-A proves the opposite: humans consistently design for the ``past war,'' optimizing for dense compute just as the world moved to sparsity. The ``Temporal Gap'' is not a hurdle for AI; it is the reason \textit{for} AI. The solution to the prediction problem is not better prophecy, but faster reaction. An automated idea factory can dramatically compress the design exploration phase---evaluating thousands of candidates in weeks rather than the months or years a human team requires. When the design cycle shrinks, the temporal gap shrinks with it; the architecture can be tuned to workloads that are current rather than forecasted. This matters more than ever: training a single frontier model now takes many months, meaning workload characteristics are increasingly stable and observable during the design window.
\section{The Automated Idea Factory}
\label{sec:idea-factory}
\label{sec:ideafactory}
We now present a concrete proposal for automated architecture exploration that can systematically discover the opportunities humans miss. The core insight is that architecture---unlike circuit layout or physical placement---is fundamentally a \emph{reasoning} problem: constrained optimization over a vast space of structural and parametric choices where the constraints are semantic (correctness, coherence, causality) rather than purely geometric. This positions it uniquely within the capabilities of modern large language models.

\subsection{System Architecture Overview}

\begin{figure}[t]
\centering
\includegraphics[width=0.45\textwidth]{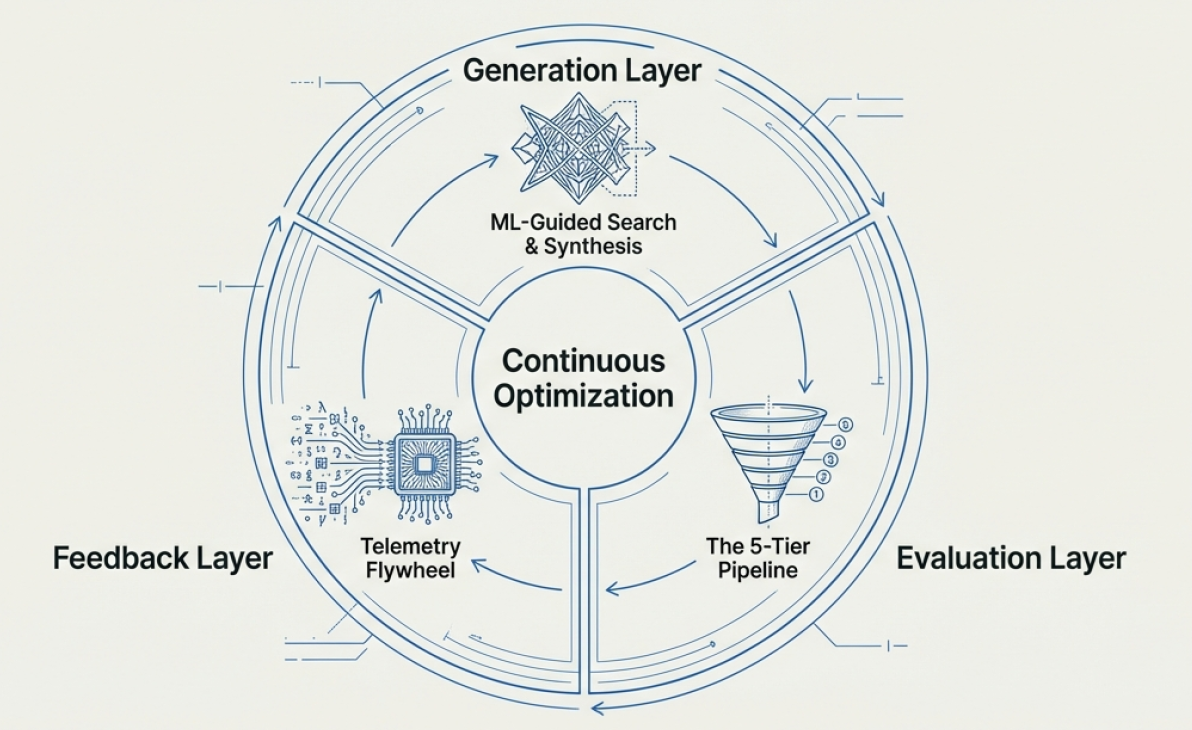}
\caption{Idea Factory Architecture}
\label{fig:idea_factory}
\end{figure}

The idea factory consists of three primary components operating in a continuous loop: a \textbf{generative discovery engine} that produces candidate architectures through reasoning-based invention rather than parameter search, an \textbf{agentic evaluation layer} that validates designs through a hierarchy spanning first-principles analysis to cycle-accurate simulation, and a \textbf{feedback layer} that collects deployment telemetry to refine problem formulation and guide future exploration. Figure~\ref{fig:idea_factory} illustrates the system architecture. The key insight is that this is not a one-shot design space exploration but a \emph{recursive} learning system that expands the frontier of architectural knowledge with each iteration---each solved problem generates new problems; each deployed chip provides telemetry that sharpens the next generation of hypotheses. Further, we emphasize that this is not a simple one-shot prompt to an LLM saying ``give me a better chip design''---it is a structured, multi-stage pipeline where each component (problem formulation, mechanism generation, adversarial critique, simulation, telemetry feedback) is purpose-built to address a specific failure mode of naive LLM application.

\subsection{Generation Layer: The Recursive Discovery Engine}

Prior work treats architectural exploration as an optimization problem amenable to parametric search, genetic algorithms, or learned surrogate models. We take a fundamentally different approach: we treat architectural invention as a \emph{reasoning} problem. The central claim is that architectural invention is deterministic given well-specified constraints---not a mysterious spark of genius, but a structured derivation from problem to mechanism. If the problem is stated precisely enough, the solution follows. This reframes the challenge: the scarce resource in architecture is not solution generation but \emph{problem formulation}---identifying the right bottleneck, the right constraint, the right question.

The generation pipeline operates in six phases:

\textbf{(1) Problem Extraction.} The pipeline begins by structuring inputs---from deployed telemetry, published literature, or human-defined constraints---into a canonical problem format specifying context, observed symptoms, and design constraints. This disciplined formulation is what enables systematic reasoning rather than open-ended brainstorming.

\textbf{(2) Mechanism Generation.} An architect agent performs root-cause analysis of the stated problem and derives candidate hardware mechanisms: specific structures (tables, buffers, state machines), policies, and datapath modifications. Critically, the agent is directed to propose \emph{mechanisms}---concrete architectural interventions---rather than incremental parameter adjustments. Each proposal includes a first-principles justification and a specification of how it should be evaluated.

\textbf{(3) Design Validation.} Generated mechanisms are assessed on two dimensions: correctness and feasibility (does the mechanism actually solve the stated problem without violating physical or logical constraints?) and novelty (is this an existing known solution, a functionally equivalent alternative, or a genuinely different approach?). Mechanisms that are incorrect or infeasible are rejected with structured feedback; those that duplicate known solutions are noted as confirmations rather than discoveries.

\textbf{(4) Recursive Problem Generation.} This phase distinguishes our approach from prior DSE. After solving a problem, the system generates \emph{new} problems through three expansion modes: Vertical---``what bottleneck emerges after this fix?'' following Amdahl's Law reasoning; Lateral---``what other domain exhibits the same structural problem?'' identifying cross-domain isomorphisms; and Foundational---``why solve this at all?'' questioning premises to redefine the design space entirely.

\textbf{(5) Divergent Exploration.} Multiple architect agents operate at different temperatures to produce a spectrum from conservative to speculative proposals for each problem, with domain-specific prompts for well-defined subproblems. This ensures the system does not prematurely converge on a single solution strategy.

\textbf{(6) Multi-Perspective Synthesis.} For high-stakes design decisions, multiple expert perspectives (microarchitecture, system integration, workload analysis, silicon feasibility) independently critique each proposal. The synthesis of these perspectives surfaces blind spots that any single viewpoint would miss.

Unlike parametric optimization which converges toward a local optimum, this recursive process \emph{expands} the frontier of exploration---the system improves by asking better questions, not just searching faster. Each solved problem spawns new problems; each failed mechanism reveals constraints that sharpen future generation. Appendix~\ref{app:generation-details} provides complete prompt specifications and worked examples.

We validate this pipeline experimentally: across 165 independent runs using a ``clean room'' protocol where the agent sees only the problem description with known solutions (from published top-tier conference papers) redacted, the pipeline achieved a 95\% success rate in producing viable mechanisms, with 32\% exact rediscoveries of known solutions and 64\% valid alternative approaches. This confirms the central claim: given a well-specified problem, architectural invention is not a bottleneck. 


\subsection{Evaluation Layer: Multi-Tier Agentic Assessment}

The evaluation layer resolves the fundamental tension between speed and accuracy through a tiered hierarchy, but with a crucial insight that inverts traditional assumptions: \textbf{quantitative evaluation is no longer the bottleneck}. Agentic systems can construct, execute, and interpret evaluations at speeds previously impossible, transforming the economics of architectural exploration.

The hierarchy spans five tiers of increasing fidelity. Tier~0 applies first-principles qualitative reasoning---does the mechanism violate causality, assume perfect prediction, or ignore critical edge cases?---filtering thousands of designs in minutes. Tier~1 subjects survivors to adversarial multi-agent analysis: Microarchitecture, Simulation Methodology, Workloads, and Systems Integration experts each score mechanisms on taxonomy dimensions, requiring consensus approval to advance. Tier~2 employs analytical models like LIMINAL~\cite{davies2025liminal}, which achieves $R^2 = 0.895$ correlation with production hardware; critically, these frameworks are trivially extensible by agentic systems---given a new mechanism, Claude can extend the model's equations, generate predictions, and identify emerging bottlenecks. Tier~3 leverages our most surprising capability: \emph{agentic simulator construction}. Given a mechanism specification, Claude Code generates purpose-built simulators modeling specific hardware structures, implements proposed policies in executable code, creates test harnesses, and interprets results. We validated this by implementing eight radically different cache policies---spanning replacement, prefetching, and partitioning with novel data structures---in hours rather than the months traditional implementation would require. Tier~4 integrates with established infrastructure (ChampSim, gem5) via agentically-generated patches and plugins. Tier~5 extends to RTL synthesis and FPGA prototyping when warranted---recent work demonstrates that agentic systems can already drive this flow autonomously from specification to tape-out-ready GDSII~\cite{designconductor2026}---though we explicitly note this is often unnecessary: for digital logic within established paradigms, the community broadly accepts cycle-accurate simulation as sufficient validation.

A typical weekly cycle processes 10,000 candidates through this funnel: 2,000 survive Tier~0, 500 pass Tier~1, 100 undergo Tier~2 analytical modeling, 20 receive purpose-built simulation at Tier~3, 5 integrate with full simulation infrastructure, and 1--2 advance toward deployment. Designs failing at any tier return to generation with structured feedback explaining the failure mode. The key claim bears emphasis: the traditional bottleneck---``implementing this would take a PhD student three months''---evaporates when implementation requires hours of agentic coding. The limiting factor shifts from ``can we evaluate this?'' to ``are we asking the right questions?''---precisely why recursive problem-generation (Phase~4) is essential. 


\subsection{Feedback Layer: Learning from Deployment}

The idea factory achieves superhuman performance through continuous learning from real-world deployment. Critically, this feedback loop does not require deploying idea-factory-designed silicon—it begins with whatever hardware is serving production workloads today. Every deployed chip, whether current-generation or legacy, already collects fine-grained telemetry: microarchitectural counters (cache rates, occupancy, stalls, power), workload characteristics (operation mix, batch sizes, access patterns), and system-level metrics (latency distributions, communication patterns, load balance). This telemetry feeds Problem Extraction, generating ground-truth problem formulations from production bottlenecks rather than synthetic benchmarks.

Deployment data calibrates every evaluation tier. Analytical models are validated against measured hardware behavior with systematic prediction errors identified and corrected. Simulation configurations are tuned to match production workload characteristics. First-principles reasoning is refined based on which failure modes actually manifested in silicon. After three years with 10,000 deployed chips, the system accumulates petabytes of workload telemetry and performance models trained on real silicon behavior---this dataset is the true moat, not architectural ideas which can be reverse-engineered. The system tracks workload evolution, clustering collected workloads and monitoring shifts over time. When MoE workloads grow from 20\% to 45\% of deployment, this idea factory automatically generates MoE-specific problems before the shift becomes a crisis. When collective communication latency emerges as the dominant bottleneck, new problem formulations target this constraint proactively. This anticipatory capability---impossible for human architects lacking systematic workload visibility---closes the loop between deployment reality and architectural exploration.

An important issue is privacy and security of deployment telemetry. Collecting detailed workload and performance data from production chips raises concerns about user privacy and potential leakage of sensitive information. We mitigate this through on-device aggregation and anonymization, ensuring that no personally identifiable information leaves the chip. Telemetry is aggregated at the datacenter level, with strict access controls and encryption to prevent unauthorized access. The system focuses on high-level performance metrics and workload characteristics rather than fine-grained traces that could reveal user behavior.

A related practical question is what telemetry infrastructure exists today. Nascent industry solutions include NVIDIA's GeForce Telemetry~\cite{geforce-experience}, and hyperscalers like Google~\cite{opentelemetry}, Meta~\cite{dynolog}, Amazon~\cite{codeguru}, Intel~\cite{intel}, and Microsoft~\cite{azure-monitor} have in-house telemetry solutions. Third-party continuous profiling services such as Datadog~\cite{datadog}, Pyroscope~\cite{pyroscope}, Parca~\cite{parca}, ydata-profiling~\cite{CLEMENTE2023126585}, and Splunk's AlwaysOn profiler~\cite{splunk} also exist. The most relevant of these is Dynolog~\cite{dynolog}, which integrates with the PyTorch profiler and Kineto CUDA profiling library. However, these solutions primarily collect software-level metrics; an orthogonal question is whether hardware performance counters alone are sufficient, or whether we need programmable introspection capabilities not envisioned in the original chip design. The latter is addressed by the recent Introspection Processing Unit work~\cite{ipu}, earlier works more than a decade ago~\cite{903267,10.1145/1168857.1168890,10.1145/1062261.1062284}.


\subsection{Why This Works Now}

The idea factory relies on technological enablers that have matured only recently. \textbf{Reasoning models} can perform the abductive reasoning required for architectural invention---our 95\% success rate demonstrates this capability is real, not aspirational. \textbf{Agentic tool use} through systems like Claude Code transforms evaluation from months of manual effort to hours of automated construction; the eight-policy implementation experiment proves this is production-ready. \textbf{Workload consolidation} around transformer architectures provides a stable optimization target---unlike the fragmented AI landscape of 2015, today's dominant workloads share common characteristics amenable to systematic optimization. \textbf{Process scaling slowdown} makes this essential: when Moore's Law delivered $2\times$ improvement automatically, architecture was secondary; now that 12nm to 3nm delivers only $2.9\times$ for AI workloads, architectural innovation is the primary lever.

Even if workload diversity increases in the future, the fundamental inflection point remains: we can now ideate, discover, evaluate, quantify, and implement architectural innovations rapidly with tireless machines. This capability---unthinkable even 12 months ago, prior to the maturity and cost-effectiveness of tools like Claude Code and OpenAI Codex---changes how we must think about architecture.  We address counterarguments, limitations, and open challenges---including evaluation fidelity, verification, paradigm shifts, and ecosystem barriers---in Section~\ref{sec:counterarguments}.

\greenbox{The question is no longer whether automated exploration can match human architects, but who will build the infrastructure to exploit it first.}

\section{Experimental Validation}
\label{sec:evaluation}
\label{sec:experiments}

Section~\ref{sec:idea-factory} described an automated idea factory comprising a recursive discovery engine, multi-tier agentic evaluation, and deployment feedback. We now present experimental evidence that each layer of this vision works today. Our experimental platform, which we call the \textbf{Gauntlet}, is a suite of multi-agent pipelines that orchestrate LLMs through structured workflows with built-in verification, adversarial critique, and ensemble selection. The Gauntlet instantiates three capabilities corresponding to the three layers of the idea factory:

\begin{enumerate}
\item \textbf{Comprehension:} Can LLMs extract deep, cross-domain architectural insights from research at a scale and depth impossible for humans? (Section~\ref{sec:eval-comprehension})
\item \textbf{Ideation:} Given only a problem description, can LLMs generate viable architectural mechanisms---and recursively expand the frontier of problems worth solving? (Section~\ref{sec:eval-ideation})
\item \textbf{Quantitative Evaluation:} Can agentic systems construct executable performance models, extend simulators, and perform design space exploration in minutes rather than months? (Section~\ref{sec:eval-quantitative})
\end{enumerate}

Taken together, these experiments substantiate the claim made in Section~\ref{sec:idea-factory}: the AlphaZero moment for computer architecture has arrived. It is available at \gauntletgitrepo{}.

\subsection{Dataset}

To validate at scale rather than on a few hand-selected examples, we assembled two datasets. The \textbf{primary corpus} comprises 85 ``mechanism'' papers---those proposing concrete architectural innovations---drawn from ISCA 2025 \fixmedone{(78 papers)} and HPCA 2026 (17 papers). All experiments used Claude Opus 4.5, whose knowledge cutoff is May 2025; aside from pre-prints on arXiv, these papers would not have appeared in the model's training data. The \textbf{curated reference set} comprises 20 established papers spanning ISCA 2009 through ISCA 2024, selected for methodological rigor and well-validated results that have stood the test of time, providing a reliable ground truth for stress-testing our quantitative evaluation pipeline.

The three experiments draw on these datasets as follows: \textbf{Comprehension} and \textbf{Ideation} used the full \fixmedone{95-paper} corpus. \textbf{Quantitative evaluation} used primarily the 20-paper curated reference set, where established results enable rigorous human verification; we also ran the pipeline against the full corpus but, due to the time required for expert review, manually verified results on a subset of 5 additional recent papers.

\subsection{Comprehension and Insight Extraction}
\label{sec:eval-comprehension}

The idea factory's evaluation layer requires that LLMs comprehend complex research across multiple domains and perspectives. We validated this with a Gauntlet comprehension pipeline that orchestrates a panel of six AI reviewers to analyze a research paper. Four reviewers are fixed domain-general personas: a microarchitecture specialist who reverse-engineers mechanisms at the bit level, a workload analyst who stress-tests evaluation methodology, a simulation tools expert who audits experimental infrastructure, and a chief architect who evaluates industry feasibility. Two additional reviewers are dynamically selected from a library of ${\sim}90$ topic-specific expert personas, matched to the paper's detected sub-topics (e.g., ``cache coherence protocols,'' ``systolic array architectures'') - topic and such expert selection also done through an agentic LLM call. Each reviewer independently produces a structured critique. A dedicated synthesizer then consumes all six reviews alongside the original paper and produces a unified ``Master Class'' reading guide---identifying points of agreement and tension across reviewers, distilling the core technical insight, and providing a frank assessment of limitations the authors downplayed.

We processed all 85 papers through this pipeline. The results easily surpassed human-generated paper insights in quality and technical depth. We highlight four representative examples below. We encourage readers to download the repository and run the Gauntlet on papers of interest; the insights are often non-obvious and reward deep technical engagement. We note that the system's critiques can be quite pointed---our goal is to demonstrate analytical capability, not to single out any particular work. The full corpus of runs is available on request and hosted here: \zeroarchpaperinsights{}.

\paragraph{Example 1: Distilling Core Mechanisms.} The Avant-Garde paper~\cite{avantgarde} sits at a deep intersection of quantization theory, numerics, and GPU hardware---with many recent advancements making expert review especially challenging. Within minutes, the system distilled the paper's contribution to its essence:

\begin{quote}
\small
Strip away the ``empowering GPUs'' language. Here's what they actually built: they added a preprocessing stage to GPUs that multiplies scaling factors into data elements before computation, plus a small modification to Tensor Cores that applies one remaining scaling factor after the dot product. That's it.

The paper addresses a real but narrow problem: when you use ``scaled numeric formats'' (like Microsoft's MX9, where groups of numbers share exponents), current GPUs can't handle the scaling natively. So you end up with ugly instruction sequences where Tensor Cores compute raw dot products, then CUDA Cores clean up by applying scaling factors element-by-element. This costs 2.14$\times$ more instructions than plain INT8.

Avant-Garde's solution: flatten multi-level scaling hierarchies into single-level ones in hardware, then let a modified Tensor Core handle the single remaining scale factor. The hardware cost is modest (claimed 1.4\% area, 1.2\% power), and the speedup is real for workloads that would otherwise suffer from the software overhead.
\end{quote}

This kind of no-nonsense distillation---cutting through marketing language to identify exactly what the hardware does and why---is precisely what architects need when surveying hundreds of papers. The full synthesis report can be obtained by running the Gauntlet on the Avant-Garde paper and is available to readers upon request, hosted here \zeroarchpaperinsights{}.

\paragraph{Example 2: Cross-Domain Insight Synthesis.}
The Lumina paper~\cite{lumina} spans graphics rendering, hardware architecture, and systems optimization---a combination that challenges human reviewers lacking expertise across all three domains. Within minutes, the system distilled the core mechanism:

\begin{quote}
\small
The entire paper rests on \textbf{one geometric insight}:

\emph{Two rays that intersect the same first $k$ significant Gaussians will produce nearly identical pixel colors.}

This works because of the sparsity they characterized in Section 2.2: only $\sim$10\% of Gaussians have transparency $\alpha > 1/255$, and 99\% of a pixel's final color comes from just 1.5\% of the Gaussians. If you match the first few significant Gaussians, you've essentially matched the ray.

\textbf{The mechanism:}
(1) During rasterization, identify the first $k=5$ Gaussians with $\alpha > 1/255$.
(2) Concatenate their IDs to form a cache tag.
(3) Look up in a 4-way set-associative cache.
(4) Cache hit: Skip remaining hundreds of Gaussian integrations, use cached RGB.
(5) Cache miss: Complete full rasterization, update cache.

\textbf{Why the hardware matters:} GPUs execute in SIMT fashion---all threads in a warp run the same instruction. When only 10\% of Gaussians are significant, 69\% of threads are masked (idle) waiting for their warp-mates. The NRU's frontend-backend split lets you filter first (all PEs check significance in parallel), then only pay for expensive color math on the $\sim$10\% that matter.
\end{quote}

This synthesis---produced in minutes---correctly identifies the geometric insight, the caching mechanism, and the hardware implications of SIMT divergence. The full synthesis report can be obtained by running the Gauntlet on the paper and is available to readers upon request, hosted here: \zeroarchpaperinsights{}.

\paragraph{Example 3: Extracting Optimization Opportunities.}
For the LIMINAL paper~\cite{davies2025liminal}, on which I am a co-author, the system identified that the gap between idealized analytical models and measured reality constitutes the optimization opportunity---an insight buried deep within the paper's validation section that human readers might skim over.

\begin{quote}
\small
\textbf{The Real Trick: Separating ``Fundamental'' from ``Implementation''}

Look at their validation (Section 5). They achieve 14.3\% MAPE, but only \emph{after} adding back real-world overheads:

\texttt{T\_Launch = 4$\mu$s} per kernel\\
\texttt{T\_Miss = 378ns} per cache miss (2 per kernel)\\
\texttt{T\_TP,Real = 10$\mu$s} per collective

This tells you exactly what the ``hardware tax'' is on current systems. The gap between their idealized model and reality is \textbf{the optimization opportunity}.
\end{quote}

This extraction---launch latency, prefetch misses, collective communication overhead---directly informs architectural insight, demonstrating how comprehension feeds the idea factory's problem formulation. The full report appears in Appendix~\ref{app:liminal}.

\paragraph{Example 4: Attention to Fine-Grained Detail.}
We inadvertently conducted a control experiment. When fed an earlier draft of the LIMINAL paper where validation relied on curve-fitting, the system noted: ``\emph{The validation isn't from first principles and doesn't provide insight into the hardware bottlenecks.}'' When fed the final version with detailed overhead modeling, it immediately extracted launch latency, prefetch misses, and collective overheads as key insights. This demonstrates that the system attends to fine-grained methodological details---not merely surface-level claims---and that this attention is critical for generating meaningful architectural insights. From our decades of experience, we assert that such fine-grained extraction across hundreds of multi-domain papers would be nearly impossible for humans at scale.

\subsection{Ideation and Design Space Exploration}
\label{sec:eval-ideation}

The idea factory's generation layer claims that architectural invention is a reasoning problem, not a spark of genius---and that the scarce resource is problem formulation, not solution generation. We validated this with the Gauntlet ideation pipeline, a four-phase protocol that extracts architectural problems from published research and tests whether LLMs can independently generate ideas and solutions.

The pipeline operates as follows. \textbf{Phase~1 (Problem Extraction):} An LLM reads only the first three pages of a published paper---enough to understand the problem but not the authors' solution---and extracts the core architectural bottleneck in a structured [CONTEXT]/[SYMPTOM]/[CONSTRAINT] format. A separate quality-control step scores symptom generality (1--10) and repairs over-specific formulations; 94\% of extracted symptoms achieved generality scores of 7 or higher. All problems were manually verified to confirm no solution leakage. \textbf{Phase~2 (Solution Generation):} A different LLM generates five independent architectural proposals per problem across a temperature sweep (0.5--0.9), ensuring diversity of approaches. \textbf{Phase~3 (Validation):} A third LLM reads the full paper and each generated solution, judging on two dimensions: similarity to the known solution (exact match, functional equivalent, or different approach) and design quality (viable or flawed). \textbf{Phase~4 (Recursive Frontier Expansion):} For each solved problem, the system generates three new problems through the Vertical, Lateral, and Foundational expansion modes described in Section~\ref{sec:idea-factory}, producing structured problem formulations ready to feed back into Phase~2.

\paragraph{Results.} The dataset comprised \fixmedone{95 papers from ISCA-2025 and HPCA-2026, yielding 475 independent runs (95 papers $\times$ 5 runs each)}. The pipeline achieved a \textbf{95\% aggregate success rate}, producing viable mechanisms in 471 of 475 attempts:

\begin{itemize}
\item In 232 runs (48\%), the agent independently re-derived the \emph{exact} mechanism proposed in the original paper---matching specific implementation details such as hashing logic, table structures, or state machine transitions.
\item In 239 runs (50\%), the agent proposed valid, high-quality mechanisms \emph{distinct} from the authors' approach, demonstrating that well-defined problems admit multiple viable architectural solutions.
\item Only 2 runs (1\%) produced flawed proposals, and all were correctly rejected by the validation layer. No single problem yielded only flawed proposals; every problem produced at least one high-quality solution.
\end{itemize}

Each solution was generated in 10--20 minutes, compared to the weeks or months a human architect would require---an acceleration of four to five orders of magnitude. The 48\% alternative success rate is particularly significant: it confirms that the scarcity in architecture lies in \emph{problem formulation}, not solution generation. Human architects, constrained by time and cognitive load, typically explore one or two solution paths per problem. The automated system explores five in under an hour, surfacing viable alternatives that human intuition would never consider. The key limiter to wall-clock time is API rate limits, since these runs are trivially parallelizable.

As a community contribution, we have also distilled the extracted problems (with solutions redacted) into an open challenge repository, providing a benchmark for evaluating both human and automated architectural reasoning. It is available here: \zeroarchproblems{}.

\subsection{Quantitative Evaluation at Unprecedented Speed}
\label{sec:eval-quantitative}

The most surprising validation concerns the evaluation tiers of our hierarchy: analytical model construction, simulator extension, and design space exploration. Traditional wisdom holds that quantitative evaluation is the bottleneck---that implementing a new mechanism in a simulator requires months of PhD-level effort. Our experiments demonstrate this assumption is obsolete.

\paragraph{Automated Performance Model Construction.}
We built a Gauntlet pipeline specifically for converting research paper claims into executable, first-principles performance models. The pipeline operates in three phases, each with its own verify-repair loop to eliminate hallucinations and ensure scientific rigor.

\textbf{Phase~1 (Text $\to$ Math):} A specification agent reads the full paper and extracts all mathematical relationships, variables, constraints, and calibration data from figures and tables into a structured specification. A verifier agent audits the specification for completeness, missing variables, and hallucinated formulas. If the verifier rejects the spec, a repair agent fixes the identified issues; the loop iterates up to three times.

\textbf{Phase~2 (Math $\to$ Code):} An implementation agent translates the approved specification into a self-contained Python model that computes both the baseline and proposed system's behavior from first principles---no hardcoded magic numbers, no external data files. Two independent verifiers run in parallel: a functional verifier checks syntax, logic, and spec fidelity, while a directive verifier enforces scientific standards (first-principles derivation, calibration against paper data, baseline comparison). The repair loop iterates until both verifiers approve.

\textbf{Phase~3 (Code $\to$ Insight):} An interpretation agent produces a human-readable explanation of the model's structure, assumptions, and findings---including any ``magic gaps'' where the paper's claimed performance exceeds what the first-principles model predicts.

To further improve robustness, we run the full pipeline three times independently and use a selector agent to pick the best run based on correctness and insight quality, producing an ensemble that is more reliable than any single run. We validated this pipeline on papers spanning transformer optimization, GPU architecture, cache management, prefetching, and branch prediction. In all cases, the system produced working, validated performance models in under 20 minutes---a process that would traditionally require weeks to months of expert human effort. Human experts reviewed the resulting models and interpretations, confirming that they were not only correct but captured the key architectural insights of the original papers. Specifically, the models produces human-understandable insights from the ISCA 2009 classic Anton paper~\cite{anton}, Darwin Genomics accelerator papers~\cite{darwin}, Google's VCU paper~\cite{vcu}, and the more recent Craterlake paper which performs FHE acceleration~\cite{craterlake}. The full set of models and insights is available on request.

We outline next two more examples of LLM-based quantitative evaluation tool construction.

\paragraph{Extending Cycle-Accurate Simulation.}
We fed the ChampSim simulation framework along with research papers describing nine different microarchitectural techniques spanning branch prediction, L1/L2 cache replacement policies, and prefetching policies. In all cases, Claude generated correct, integrated implementations in 10--20 minutes. Each implementation included novel data structures, integration with the simulation framework's APIs, and functional validation. This confirms that agentic systems can extend established simulation infrastructure as readily as they construct new models.

\paragraph{Extending Analytical Models.}
Given the LIMINAL analytical framework~\cite{davies2025liminal}, Claude produced working code for the model in minutes. More significantly, given hyperparameters for Nemotron (a model using Mamba2 layers, which are fundamentally different from the attention, FFN, and MoE layers already modeled) and the SSM paper~\cite{dao2024transformersssmsgeneralizedmodels}, Claude generated a correct analytical extension with no human coding required. This is particularly powerful because Mamba2's state-space formulation differs architecturally from transformer attention---yet the system reasoned about the new computational patterns and produced accurate predictions. This validates our claim that analytical frameworks are trivially extensible by agentic systems---a critical enabler for rapid design space exploration.

\subsection{Summary}

These experiments validate the core claims of Section~\ref{sec:idea-factory}. Multi-perspective comprehension extracts insights across domains that human specialists would miss, enabling the cross-domain reasoning that drives architectural innovation. LLM-based ideation achieves 95\% success in generating viable mechanisms, with 64\% producing valid alternatives to published solutions---confirming that problem formulation, not solution generation, is the scarce resource. Quantitative evaluation, traditionally the bottleneck, now takes minutes through agentic tool construction---shifting the constraint from ``can we evaluate?'' to ``what should we explore?''
\greenbox{The frontier of architectural innovation can now move faster than human design cycles.}



\section{Counterarguments and Limitations}
\label{sec:counterarguments}

We now address potential criticisms of our proposal and acknowledge genuine limitations. The counterarguments we consider are serious and deserve careful response; dismissing them would undermine the credibility of our thesis.

\subsection{Paradigm Shifts and Local Optima}\label{sec:paradigm-shifts}

The critique: automated search might get trapped in local optima, finding incrementally better designs within a paradigm but missing the paradigm shifts that drive transformative progress---transitions as fundamental as Von Neumann to dataflow, in-order to out-of-order execution, or GPU to TPU design philosophies. Human creativity, the argument goes, is essential for such leaps. The arguments advanced in recent works like ``LLMs can't jump''~\cite{zahavy2026llms} reinforce this concern. 

We take a strong contrary position: \textbf{reasoning models are more likely to discover paradigm shifts than human architects}, precisely because of capabilities humans lack.

Consider how historical paradigm shifts actually occurred. Dataflow architectures emerged from functional programming and lambda calculus. Out-of-order execution borrowed from compiler instruction scheduling. Systolic arrays applied signal processing concepts to computation. SIMT drew from graphics pipeline parallelism. In each case, the breakthrough came from \emph{cross-domain transfer}---applying concepts from one field to architectural problems in another. Human architects, trained in narrow specializations, rarely possess the breadth to make such connections systematically. The ``insight'' attributed to human designers is often post-hoc rationalization of what the constraints demanded: RISC arose as a response to compiler technology and VLSI constraints; out-of-order execution emerged to hide memory latency.

Our recursive problem generation pipeline is explicitly designed for cross-domain discovery. Its Lateral expansion mode identifies abstract structural isomorphisms across domains, surfacing connections that domain-specialist humans would miss. Its Foundational expansion mode goes further, questioning whether the problem should be solved at all---the prerequisite for paradigm-level rethinking. The design space our system can explore is bounded only by digital CMOS physics, whose rules the underlying foundation models understand from extensive training on hardware documentation, physics texts, and engineering literature. Within this space, the system can reason about:

\begin{itemize}
    \item \textbf{Information theory:} Shannon limits, entropy bounds, and minimum-energy computation principles that constrain what architectures can achieve
    \item \textbf{Sampling theory:} Nyquist-style reasoning about when computation can be approximated, compressed, or skipped entirely
    \item \textbf{Control theory:} Feedback, stability, and adaptive systems that could inform self-tuning architectures
    \item \textbf{Queueing theory:} Little's Law and scheduling principles that govern resource allocation under contention
    \item \textbf{Statistical mechanics:} Phase transitions and emergent behavior in large-scale parallel systems
    \item \textbf{Game theory:} Nash equilibria and mechanism design for multi-agent resource allocation
    \item \textbf{Category theory:} Compositionality and abstraction principles for modular hardware design
    \item \textbf{Coding theory:} Error correction and redundancy trade-offs for reliability under variation
\end{itemize}

Each of these fields contains insights that could fundamentally reshape architectural thinking---but few human architects have deep expertise across all of them. A reasoning model trained on the corpus of human knowledge can draw connections that no individual specialist would make. Furthermore, AlphaGo and AlphaZero discovered paradigm shifts in their domains~\cite{silver2017alphazero}---opening moves no human had considered, positional evaluations that upended centuries of chess theory. If automated search can discover such shifts in Go and chess, there is no principled reason it cannot do so in computer architecture.

A genuine concern remains: radically revolutionary paradigms like quantum computing or neuromorphic computing may require physical insights that reasoning within classical digital logic cannot discover. However, such revolutions are rare---decades pass between them. Within a technology paradigm, automated exploration should dominate. We therefore view paradigm discovery not as a limitation but as a \emph{strength} of the reasoning-based approach---one that will compound as foundation models improve and as the recursive exploration accumulates insights across domains.


\subsection{Verification and Validation}

The critique: machine-generated architectures might harbor subtle bugs or corner cases that human designers would catch through intuition and experience. Silicon bugs are catastrophic---you cannot ship chips with bugs, and the cost of a respin can be prohibitively expensive.

Our response has several layers. First, compositional generation from verified building blocks substantially reduces the verification burden. Architectures are constructed by composing pre-verified components---standard cache controllers, proven execution core designs, established interconnect protocols. Novel functionality is limited to interfaces between known-good components.

Second, formal verification tools are increasingly powerful~\cite{kaufmann2018formal}, including automated theorem proving, model checking for finite-state systems, and equivalence checking between RTL and specification. Machine-generated designs may actually be \textit{more} verifiable than human designs because they tend to be more regular and compositional, with fewer clever-but-fragile optimizations that complicate formal analysis.

Third, a subtler advantage emerges from the reasoning-based approach: generated mechanisms come with explicit first-principles justifications that make assumptions transparent and auditable. Traditional optimization produces designs that work without explaining why---making failure mode analysis difficult. Our approach produces designs with causal reasoning chains that verification engineers can scrutinize directly.

Fourth, extensive FPGA prototyping with real workloads catches functional bugs, corner cases under stress, timing violations, and power and thermal issues before tapeout. By the time a design reaches silicon, it has been validated at scale with production-representative workloads.

Finally, humans make bugs too. The Intel Pentium FDIV bug~\cite{intel1994pentium}, AMD TLB errata, and Apple M1 speculative execution issues demonstrate that human-designed architectures have bugs despite enormous validation efforts. There is no evidence that humans are systematically better at avoiding bugs; both approaches require rigorous verification.

The remaining concern is that genuinely novel architectural features may have unknown failure modes that neither humans nor automated systems anticipate. Mitigations include introducing novelty incrementally (one new feature per generation) and small initial production runs before volume manufacturing.

\subsection{The Evaluation Fidelity Gap}

The critique: simulation is never perfect. Designs that look great in simulation often underperform in silicon due to manufacturing variation, workload differences between benchmarks and production, power modeling errors, thermal effects, and microarchitectural interactions not captured in models. If evaluation is inaccurate, automated search optimizes the wrong objective---Goodhart's Law~\cite{goodhart1984} applied to architecture.

This critique is valid, and it is precisely why deployment feedback is central to our approach. The feedback loop addresses the fidelity gap directly: deploy chips with telemetry, measure actual performance against predictions, identify systematic errors, calibrate models, and repeat. Over time, prediction accuracy improves because models learn from ground truth.

The agentic evaluation paradigm also provides new mitigations unavailable to traditional approaches. First-principles reasoning at the earliest evaluation tiers catches designs that violate physical constraints before simulation is even attempted. Multi-agent adversarial review surfaces methodological concerns that pure numerical optimization would miss. The multi-tier pipeline reduces risk by catching issues early at appropriate fidelity levels---only designs that survive increasingly rigorous scrutiny advance toward silicon.

The system can also reason probabilistically about uncertainty: ``This design predicts 2.3X $\pm$ 0.3X improvement with 90\% confidence'' or ``Manufacturing variation could reduce performance by 15\% in the worst case.'' Uncertainty-aware optimization prevents overfitting to unreliable predictions.

The remaining concern is that first-generation designs lack deployment feedback, so prediction errors will be larger. Mitigations include conservative targets (aim for 2X improvement; delivering 1.5X is still success) and extensive FPGA validation before tapeout.

\subsection{Objective Function Misspecification}

The critique: the system optimizes what you tell it to optimize. If the objective function is misspecified---for example, optimizing MLPERF scores that do not reflect real usage---you get designs that game the metric without delivering actual value.

Multi-objective optimization provides the first defense. We optimize throughput, latency, power efficiency, and cost simultaneously. Gaming one metric typically degrades others; a design must perform well across all objectives to rank highly. Evaluation uses diverse real workloads from telemetry spanning different model sizes, architectures, tasks, and deployment scenarios. A design that performs well across this diversity is genuinely good, not gaming a narrow benchmark.

Deployment provides ultimate validation. Do customers actually use the chips? Do they achieve expected cost savings? If optimized objectives diverge from real value, telemetry reveals the mismatch and objectives are refined. The recursive problem-generation capability also helps: by continuously reformulating problems based on production bottlenecks, the system naturally gravitates toward objectives that matter in practice.

The remaining concern is that the initial objective function will inevitably be imperfect. Mitigations include starting with conservative, proven metrics, iterating based on deployment experience, and maintaining human oversight for major decisions like tapeout go/no-go.

\subsection{Manufacturing, Supply Chain, and Software Ecosystem}

The critique: even if you design a better chip, bringing it to market is hard. On the hardware side, leading foundries prioritize their largest customers for capacity allocation, supply chains for HBM, advanced packaging, and testing are complex, and manufacturing yield issues can torpedo economics. On the software side, any new architecture requires a software stack---compilers, libraries, frameworks---and the cost of building and maintaining this stack has historically been a major barrier to entry.

As discussed in Section~\ref{sec:implications}, the use of trailing-edge nodes (12nm or 7nm rather than 3nm) directly addresses hardware supply chain concerns. These nodes have ample capacity at multiple foundries, are cheaper, and have mature yields with fewer manufacturing surprises. Early engagement with fab partners during the design phase enables design-for-manufacturability review and advance capacity reservation. Using standard components for non-core elements---standard HBM, standard packaging, standard testing infrastructure---limits novelty to the die design itself, reducing supply chain complexity.

On the software side, the barrier is lower than it has ever been, for two reasons. First, existing startups have already demonstrated that new architectures can achieve software ecosystem viability, particularly for inference workloads where high-level frameworks like PyTorch abstract hardware details. The ecosystem lock-in concern, while real, is empirically surmountable. Second, and more fundamentally, the same AI-driven productivity revolution that enables the idea factory also transforms software development. Agentic coding tools can generate compilers, kernels, and optimization passes at speeds unimaginable even two years ago (for kernels for GPUs exxamples include VibeTensor~\cite{xu2026vibetensorsoftwaredeeplearning} and Makora~\cite{makora}). The software stack that once required a team of 50 engineers over multiple years can increasingly be built and maintained by a smaller team augmented with AI coding assistants. This is not speculation---it is the same capability shift that underpins the idea factory itself.

The remaining concern is that both manufacturing and software ecosystem development are genuinely hard, and new entrants have failed at both despite good designs. Mitigations include hiring experienced experts in both domains, starting with small volumes, and targeting deployment scenarios where the software burden is manageable.

\subsection{Genuine Limitations}

Beyond counterarguments that we believe are addressable, we acknowledge genuine limitations that represent real constraints on our approach.

The first limitation is \textbf{first-generation risk}. First-generation chips lack deployment feedback, so prediction accuracy will be lower than for subsequent generations. If the initial chip underperforms expectations, organizational commitment may waver before the second iteration can demonstrate the compounding improvement that the feedback loop enables. Conservative targets and extensive validation provide partial mitigation, but the risk is real.

The second limitation is \textbf{team and organizational challenges}. Building the idea factory requires a rare combination of skills: computer architecture expertise, machine learning systems knowledge, silicon design and verification capability, and manufacturing and operations experience. Moreover, the organizational culture must embrace data-driven decision making, embracing AI recommendations over human intuition, and fast iteration with acceptance of failure. This represents a significant culture shift from traditional chip companies.

A final consideration is \textbf{workload evolution}. If AI workloads change dramatically---for example, a paradigm shift beyond transformers---any architecture optimized for the current regime could become suboptimal. However, this concern actually strengthens the case for the idea factory rather than weakening it. Whether workloads ossify around a stable paradigm or explode in variety, the need for rapid architectural adaptation remains---and an automated system that can ideate, evaluate, and iterate in weeks will always outpace a human team operating on multi-year design cycles. As argued in Section~\ref{sec:exceptionalism}, the solution to the prediction problem is not better prophecy but faster reaction. Workload volatility is a threat to \emph{slow} design processes; it is an opportunity for fast ones.


\section{Implications}
\label{sec:implications}

The idea factory is not merely a research proposal---it is an inflection point whose
consequences differ sharply depending on where an organization sits in the semiconductor
ecosystem. Below, we first distill two technical observations from the preceding sections,
then examine how automated architecture redefines the mandate for academic research and reshapes the industry's structure.

\paragraph{Observation 1: Architecture now dominates process.}
Davies and Sankaralingam~\cite{davies2025defying} show that technology scaling from 12nm
to 3nm delivers at most 2.9$\times$ improvement for transformer workloads---far below
the 8$\times$ classical Moore's Law would predict over three node generations.
Meanwhile, architectural specialization at 12nm can exceed the performance of
general-purpose designs at 7nm by 2.5$\times$ on inference and 2.2$\times$ on training,
with 7$\times$ better power efficiency. A corollary is that trailing-edge nodes become
viable platforms for competitive designs: automated exploration on 12nm, with its
3--4 month tapeout cycles, can iterate far faster than manual design on 3nm with
6--9 month cycles. The implication is stark: the returns to architectural innovation
now exceed the returns to process scaling.

\paragraph{Observation 2: Evaluation speed is the new bottleneck---and it has been
broken.}
Section~\ref{sec:experiments} demonstrated that agentic tool construction reduces
mechanism implementation from months to minutes. Combined with multi-tier evaluation
that filters 10,000 candidates per week down to 1--2 deployment-ready designs, the
binding constraint shifts from \emph{``can we evaluate this?''} to \emph{``what should we
explore?''} The team with the fastest, most accurate evaluation pipeline---calibrated
by deployment telemetry---holds the decisive advantage. This is a \emph{data} advantage
and an \emph{infrastructure} advantage, not an \emph{ideas} advantage. Ideas, as our
95\% success rate shows, are abundant.

\subsection{Implications for Research Labs and Academia}
If our thesis is correct, the nature of computer architecture research changes
fundamentally. Three shifts deserve attention.
\begin{itemize}

\item \textbf{From solutions to problems.} Our 95\% success rate on solution generation,
combined with 64\% producing valid alternatives to published solutions, confirms that
\emph{problem formulation}---not solution creativity---is the scarce resource. Research
impact will increasingly depend on identifying the right problems: defining objectives,
formalizing constraints, and specifying what ``better'' means for a given context.
Papers whose primary contribution is a clever mechanism for a well-defined problem will
face diminishing returns as automated systems generate such mechanisms routinely.

\item \textbf{From architectures to evaluation methodology.} If automated exploration
generates thousands of candidate architectures, the binding constraint becomes evaluation
quality: accuracy, speed, and calibration against real hardware. Research on faster
simulation, more accurate learned performance models, uncertainty quantification, and
privacy-preserving telemetry collection becomes the critical path. We expect evaluation
methodology to become the most impactful subfield of computer architecture research
within a decade.

\item \textbf{From closed benchmarks to open telemetry.} The architecture community's
reliance on synthetic benchmarks (SPEC, MLPerf) is a known limitation~\cite{hoefler2015scientific}.
The idea factory's dependence on deployment telemetry makes this limitation acute: models
calibrated on synthetic workloads will systematically mispredict production behavior. We
urge the community to develop shared, privacy-preserving telemetry infrastructure---analogous
to genomics data repositories---that enables calibrated evaluation without compromising
proprietary workload details. Without this, the gap between industrial and academic
architecture research will widen irreversibly.
\end{itemize}

\subsection{Implications for industry}

\paragraph{The Three Pillars of Automated Architecture}
The idea factory's dependence on closed-loop feedback---deploy, measure,
recalibrate---creates a structural reorganization of the semiconductor ecosystem. Success
in this paradigm requires integrating three distinct capabilities: (1)~\emph{deployment
telemetry} to calibrate evaluation models, (2)~\emph{agile cloud infrastructure} to
rapidly deploy and test candidates, and (3)~deep \emph{silicon verification legacy} to
ensure machine-generated designs are physically manufacturable. The industry is converging
toward this triad from two opposing directions.

\paragraph{The Top-Down Path: Vertically Integrated Providers}
Vertically integrated organizations---cloud providers with custom silicon programs---approach
this from the top down. They natively control the infrastructure and operate millions of
accelerators running production workloads, granting them exclusive access to the critical
deployment telemetry required to close the automated evaluation loop. Their structural
challenge, however, lies in silicon legacy: their internal hardware teams are historically
much smaller than those of traditional chipmakers. For these organizations, the idea
factory is a mechanism to amplify their small silicon teams, allowing them to iterate
thousands of candidate architectures in simulation to bypass their disadvantage in
historical design expertise.

\paragraph{The Bottom-Up Path: Merchant Silicon and Neoclouds}
Conversely, merchant silicon providers possess unmatched verification infrastructure,
historical design datasets, and manufacturing relationships---the ideal foundation for
generative design, treating accumulated architectural expertise as \emph{training data}
for generative models, verification suites as \emph{validators} for machine-generated
designs, and manufacturing relationships as \emph{enablers} of rapid iteration.
Historically, their critical weakness was a lack of access to closed-loop deployment
telemetry, as they sat behind hyperscalers' proprietary serving stacks. However, these
providers are actively closing this gap through the emergence of \emph{neoclouds}---tightly
aligned cloud providers built exclusively on merchant hardware---and direct partnerships
with frontier model providers. By investing in dedicated, proxy cloud infrastructure,
merchant silicon vendors are constructing an alternative vertically integrated stack that
secures access to production telemetry and rapid deployment capabilities without relying on
traditional hyperscalers.

\paragraph{Convergence}
The implication is that the traditional boundary between ``chip company'' and ``cloud
provider''---a line already blurring due to the capital demands of AI
infrastructure---will fully dissolve under the pressure of automated architectural
exploration. To push computer architecture past the current limits of human-driven
scaling, organizations must seamlessly connect decades of verification legacy (to ground
the generative models) with live deployment telemetry (to calibrate the objective
function). Section~\ref{sec:predictions} expands on this with specific predictions about
GPU programmability's logical endpoint.

\section{Predictions and Future Directions}
\label{sec:predictions}

We offer specific, falsifiable predictions about how computer architecture will evolve if our thesis is correct. These predictions provide concrete tests of our arguments and suggest directions for future research. We organize predictions into two categories: \emph{how architecture will be practiced} (process and methodology changes) and \emph{what automated architecture will deliver} (concrete artifacts and outcomes). To avoid the ambiguity of qualitative language (e.g., ``likely'' or ``highly probable''), we assign explicit percentage probabilities to each prediction. This aligns with rigorous forecasting methodologies and allows the architecture community to retrospectively score our accuracy in five years.

\subsection{How Architecture Will Be Practiced}

The following predictions address changes in methodology, tooling, and organizational structure. We assign these higher confidence because the underlying trends---ML adoption in design workflows, maturation of agentic tools, and organizational pressure to reduce design cycles---are already observable.

First, computer architecture job postings at major chip companies will increasingly emphasize ML and data science skills over traditional architecture expertise, with job titles like ``Architecture Machine Learning Engineer'' becoming common. Evidence would appear in job postings from leading semiconductor and hyperscalar companies showing ML, Python, and learned-model requirements in architecture team listings, contrasting with historical emphasis on RTL, Verilog, and circuit design. We assign 75\% confidence; even if full automation does not materialize, ML-aided design will grow.

Second, within two years more than 50\% of new computer architecture papers will incorporate ML-based design space exploration or learned performance models as the core ideation technique instead of the ``human aha''. Evidence would come from surveying major conferences (ISCA, MICRO, HPCA) and counting papers using automated search, neural architecture search, or ML-based evaluation. Depending on ACM policies on disclosure/banning of AI for ideation, researchers may never disclose this. We assign 70\% confidence; the research community is conservative, but evidence supporting these methods is mounting.

Third, within two years, at least one paper will demonstrate end-to-end automated chip design from workload specification to tapeout with minimal human intervention---humans setting only objectives and constraints---where the resulting architecture embodies a \emph{novel} idea not previously known. Recent work has already demonstrated autonomous agents that implement known architectures end-to-end: Design Conductor~\cite{designconductor2026} produces a tape-out-ready RISC-V core from a brief requirements document in 12 hours. Our prediction is farther along the creativity spectrum: not automating the \emph{implementation} of a conventional design, but automating the \emph{invention} of an unconventional one. Evidence would include academic or industry research showing full automation through ideation, generation, simulation, synthesis, and tapeout, with human roles limited to high-level goals. We assign 65\% confidence; the technical challenges are surmountable but require dedicated effort.

Fourth, the majority of computer architecture research papers will address evaluation methodologies, objectives, and constraints rather than proposing specific architectural ideas. The rationale: if automated search dominates design exploration, human contribution shifts to developing better evaluation methods (more accurate, faster), better objective functions (what should we optimize?), and better constraint specifications (what is feasible?). Proposing specific architectures becomes pointless when machines explore the space more thoroughly. We assign 60\% confidence; this is speculative but follows logically from our thesis.

Fifth, at least one major chip company will eliminate traditional architecture teams in favor of ``architecture AI'' groups focused on building and maintaining automated exploration systems. Evidence would include organizational restructuring announcements, transitions from traditional architects to ML engineers, and public statements about AI-driven design. We assign 50\% confidence; companies are conservative about eliminating proven processes.

Sixth, IP licensors will pivot from selling pre-designed cores to offering ``architecture-as-a-service.'' Today, the value of a major IP licensor lies in proven, verified, software-compatible cores that customers integrate into their designs. But if automated architecture generation commoditizes core design itself, this value erodes. What remains valuable is the infrastructure: decades of verification test suites, formal methods expertise, manufacturing relationships, and the software ecosystem built on established ISA binaries. We predict that leading IP companies will offer a new model---customers provide workload telemetry, and the licensor's idea factory generates a custom SoC, validates it against silicon-proven verification infrastructure, and delivers a tapeout-ready design. The moat shifts from ``we have the best cores'' to ``we have the best process for generating and validating cores.'' We assign 50\% confidence; the business model transformation is significant, but the alternative---watching IP value erode as generation is commoditized---is worse.

\subsection{What Automated Architecture Will Deliver}

The following predictions address concrete artifacts and performance outcomes. We assign these lower confidence because outcomes depend on execution quality, workload evolution, and market dynamics---factors less predictable than methodological trends.

First, the best-in-class AI inference accelerators measured by performance per dollar will come from companies using ML-guided automated design rather than traditional human-driven architecture teams. The test: measure performance per dollar on a standard inference benchmark suite with diverse models and realistic workloads, comparing startups using automated design against incumbents using traditional methods. We assign 65\% confidence; this is the core thesis, with moderate confidence reflecting execution risk.

Second, the performance-per-dollar ratio for automated versus human-designed accelerators will exceed 2X on realistic AI inference workloads. The test: take the best automated design (from startup or research project) and the best conventionally designed accelerator, measure on diverse inference workloads, and verify that automated designs win by at least 2X. We assign 55\% confidence; this is ambitious and depends on both execution quality and actual adoption of automated approaches.

Third, autonomous vehicle and robotics companies will treat compute hardware as a field-upgradeable module on an 18--24 month refresh cycle, with fleet telemetry as the first-class input to architecture generation. Several AV companies already perform hardware retrofits across deployed fleets; the pattern will become standard across the industry. The rationale is economic: algorithmic improvement in autonomy is so rapid that compute hardware becomes obsolete before the vehicle's mechanical life ends, yet a \$500 board upgrade is far cheaper than vehicle replacement. We predict standardized ``compute bays'' that accept modular boards, with the telemetry $\rightarrow$ design $\rightarrow$ deploy cycle running continuously. Fleet operation reveals perception bottlenecks; the idea factory generates optimized next-generation architecture; new boards ship to service centers; vehicles receive hardware upgrades like phones receive software updates. This compresses the hardware/software co-evolution cycle from 5-year generations to 18 months, with the telemetry moat compounding---more miles driven yields better architecture yields better autonomy yields more customers yields more miles. We assign 45\% confidence; the logistics of field upgrades are complex, but the economic pressure toward continuous hardware improvement is substantial.

\begin{table}[t]
\centering
\caption{The evolution of GPU programmability. Each transition occurred when the design space became too large to anticipate all uses with fixed hardware. The idea factory represents the next step: generality at the design process itself.}
\label{tab:generality-evolution}
\begin{tabular}{lll}
\toprule
\textbf{Era} & \textbf{What's Fixed} & \textbf{What's Programmable} \\
\midrule
Fixed-function OpenGL & Entire pipeline & Nothing \\
Pixel shaders (2001) & Vertex, rasterization & Fragment color \\
Vertex shaders (2002) & Rasterization, fixed stages & Geometry + fragments \\
Unified shaders (2006) & Pipeline structure & Any stage, any shader \\
GPGPU/CUDA (2007) & Hardware architecture & Arbitrary parallel compute \\
Tensor cores (2017) & Core operations & Matrix dims, data types \\
\midrule
\textbf{Idea Factory (202X)} & \textbf{Manufacturing physics} & \textbf{The architecture itself} \\
\bottomrule
\end{tabular}
\end{table}

Fourth, the accelerator industry's two-decade evolution toward generality will reach its logical conclusion: programmable architecture generation. GPUs have steadily moved the boundary of programmability upward (Table~\ref{tab:generality-evolution}). Fixed-function OpenGL pipelines gave way to pixel shaders, then vertex shaders, then unified shader architectures, then general-purpose compute models, then programmable tensor operations. Each transition occurred when the design space became too large to anticipate all uses with fixed hardware. The idea factory represents the next---and perhaps final---step in this progression: rather than building flexible hardware that can be programmed after fabrication, we build a flexible \emph{design process} that generates purpose-optimized hardware before fabrication. Generality moves from runtime to design time. Incumbent accelerator vendors are uniquely positioned to lead this transition; their decades of architecture expertise becomes training data for generation models, their verification and tooling infrastructure validates generated designs, and their manufacturing relationships produce the resulting silicon. The vendor that recognizes this as evolution---not disruption---and invests in idea factory infrastructure will define the next era of accelerator computing. We assign 50\% confidence that this framing becomes explicit in industry roadmaps within five years; the technical trajectory is clear, but organizational inertia may delay recognition.

Our predictions are deliberately specific and falsifiable. By 2030, we will know whether automated exploration has transformed the practice of computer architecture or whether our thesis overestimates the readiness of the technology. In either case, the evidence presented in this paper---a 95\% success rate on mechanism generation, orders-of-magnitude speedups in evaluation, and systematic documentation of human design failures---demands that the community take the question seriously.

\section{Related Work}
\label{sec:related}

\paragraph{Neural Architecture Search as Precedent}
Neural Architecture Search (NAS)~\cite{zoph2016neural,elsken2019neural} provides the
closest precedent for our thesis: automated search outperforming human experts in complex
design spaces. NAS-discovered neural network architectures now match or exceed
hand-designed models across vision and language tasks~\cite{zoph2018learning,tan2019efficientnet},
using reinforcement learning~\cite{zoph2016neural}, evolutionary methods~\cite{real2019regularized},
and gradient-based optimization~\cite{liu2018darts}. Our work extends this principle from
software architectures (computation graphs) to hardware architectures (physical silicon
with area, power, and manufacturability constraints), with a fundamentally different
evaluation cost structure that motivates multi-tiered filtering rather than end-to-end
training.

\paragraph{ML in Architecture: From Parameter Tuning to Structural Synthesis}
Machine learning has been applied to hardware design space exploration for over a
decade---learning performance models from simulation
data~\cite{ipek2006efficient,lee2007methods}, Bayesian optimization over cache and
pipeline parameters~\cite{zuluaga2016varepsilon}, and reinforcement learning for
component-level tuning~\cite{janapa2023archgym}. These approaches share a common
limitation: they optimize \emph{parameters within a human-defined structural paradigm}.
The architect chooses the memory hierarchy topology, the execution model, and the
interconnect strategy; ML tunes knobs within that fixed frame. Our work operates at a
different level of abstraction: generating the structural paradigms themselves---new
execution models, novel memory organizations, unconventional dataflow strategies---then
evaluating them through the same ML-accelerated pipelines that prior work developed for
parameter search.

\paragraph{From Stochastic Search to Deterministic Reasoning}
Traditional design space exploration~\cite{palermo2005respect,10.1145/1815961.1815968},
including ML-guided variants, treats architecture as a stochastic optimization
problem: sample the space, evaluate candidates, update a surrogate model, repeat. Even
recent frameworks like ArchGym~\cite{janapa2023archgym} and BOOM
Explorer~\cite{10.1145/3630013} operate within this paradigm, achieving incremental
improvements over baselines through more efficient sampling. Our idea factory treats
architecture as a \emph{deterministic reasoning and problem-formulation} task: given a
workload characterization and a set of hardware constraints, synthesize a mechanistic
model, derive architectural implications, and generate candidate designs with explicit
rationales. This is not a better optimizer---it is a different category of
system, closer to an automated researcher than an automated search algorithm. Benchmarks
such as QuArch~\cite{prakash2025quarch}, which evaluate architectural knowledge through
multiple-choice recall, measure a prerequisite skill that modern reasoning models have
already surpassed; the frontier has moved from passive recall to generative synthesis.

\paragraph{Agentic AI Systems Beyond Hardware}
Our approach reflects a broader trend: the application of LLM-based agentic systems to
domains previously requiring deep specialist expertise. In software engineering, agents
autonomously resolve real-world GitHub issues~\cite{jimenez2024swebench}. In chemistry,
LLM agents plan and execute synthesis experiments~\cite{boiko2023autonomous,bran2024chemcrow}.
In mathematics, program-search agents have discovered constructions surpassing the
best-known human results~\cite{romera2024funsearch,alphaevolve_agent}. Most strikingly,
Lu et al.~\cite{lu2026aiscientist} demonstrate end-to-end automation of the AI research
cycle itself---from hypothesis generation through experimentation to manuscript
production---establishing that the pattern extends to research as a whole, not merely to
narrow subtasks. We adapt this paradigm to the specific constraints of silicon
hardware---where evaluation is orders of magnitude more expensive than in software,
designs must satisfy hard physical constraints, and the feedback loop spans months from
tapeout to deployment telemetry. The idea factory is thus both a contribution to computer
architecture and an instance of the general pattern: domain expertise, once thought
irreducibly human, yields to systematic LLM-driven exploration when paired with
appropriate evaluation infrastructure.

\section{Conclusion}
\label{sec:conclusion}

The end of Moore's Law has shifted the primary lever for performance improvement from process scaling to architectural innovation, yet the architectural design space is incomprehensibly larger than human teams can explore. Through retrospective analysis and experimental validation, we demonstrated that LLM-based reasoning systems can generate publication-worthy architectural mechanisms with 95\% success, surface valid alternatives that human authors never considered in 64\% of cases, and enable quantitative evaluation at speeds that compress months of implementation effort into minutes. The scarcity in computer architecture is not solutions---it is problems.

These capabilities make the automated idea factory---a continuous loop of generation, multi-tier evaluation, and deployment feedback---not only feasible but inevitable. The decisive competitive advantage shifts from process node access and accumulated human intuition to evaluation infrastructure and proprietary telemetry data. This transition affects every constituency in the ecosystem: incumbents face the innovator's dilemma, hyperscalers are best positioned to act, startups can exploit trailing-edge architectural arbitrage, and the research community must pivot from proposing mechanisms to building the evaluation methodology and open telemetry infrastructure that automated exploration demands.

Computer architecture will not disappear---it will transform. Human architects will shift to higher levels of abstraction: defining objectives, specifying constraints, and formulating the problems that machines solve. The core activity of exploring vast design spaces to discover high-performance architectures has reached its AlphaZero moment. The mandate for the research community and industry is no longer to compete on human intuition, but to build the closed-loop evaluation infrastructure capable of pushing hardware scaling into its next era.

The principles demonstrated here---large search spaces, evaluable objectives, automated exploration---are not unique to computer architecture. Compiler optimization~\cite{ansel2014opentuner,chen2018tvm,xu2026vibetensorsoftwaredeeplearning,makora}, materials science~\cite{curtarolo2013high,butler2018machine}, and drug discovery~\cite{senior2020improved} face structurally identical challenges, and we expect similar automated idea factory approaches to emerge across engineering domains where resources no longer improve automatically.

\section*{Acknowledgments}
This work benefited from discussions with several researchers in the computer architecture and machine learning systems communities. We specifically wish to thank Jenny Huang, Chinnikrishna Kothapalli, Hrishikesh M S, and Vikas Singh who provided feedback on very early drafts of the paper. Thanks also to Angshuman Parashar, Michael Pellauer, Christos Kozyrakis, Benjamin Klenk, Oreste Villa, Tej Chajed, Sandeep Silwal, and Ming Liu for feedback.

\bibliographystyle{IEEEtran}
\bibliography{references,references_penrose}
\clearpage\newpage
\appendix
\section{Generation and Evaluation Pipeline Details}
\label{app:generation-details}
\label{app:evaluation-details}

This appendix provides complete specifications for the generation and evaluation pipelines described in Section~\ref{sec:idea-factory}, including prompt templates, taxonomy definitions, and implementation statistics.

\subsection{Problem Extraction Format}

All problems are structured in a canonical format ensuring the generation phase receives well-specified inputs amenable to systematic reasoning:

\begin{itemize}
    \item \textbf{[CONTEXT]:} The system configuration and workload characteristics under which the problem manifests.
    \item \textbf{[SYMPTOM]:} The observed performance bottleneck with quantitative evidence (utilization rates, latency breakdowns, throughput limits).
    \item \textbf{[CONSTRAINT]:} What prevents naive solutions from working---the reason this problem is hard.
\end{itemize}

Problems are extracted from the published Literature. We process the first three pages of architecture papers to extract problem statements while strictly redacting proposed solutions. This creates a corpus of validated problems with known solutions for calibration and validation. GeminiPro2.5 was used for problem extraction, with another separate prompt check the extracted problem to confirm no solution leakage.

\subsection{Phase 2: Architect Agent Prompt Template}

The Architect Agent operates with the following prompt structure shown in Figure~\ref{fig:architect-prompt}

\begin{figure*}[t]
\begin{verbatim}
[EXPERIMENTAL CONTEXT]
You are an AI agent testing "Automated Architectural Invention." 
You are a Distinguished Researcher aiming for an ISCA/MICRO paper.
You are receiving a problem description from a "Clean Room."

[YOUR TASK]
1. Analyze the root cause.
2. Propose a NOVEL hardware micro-architecture mechanism to solve it.
   - Do NOT propose incremental tuning.
   - Be specific about hardware structures (tables, buffers, logic).
3. Outline the experimental design.

[PERFORMANCE REPORT]
{symptom_report}

[OUTPUT REQUIREMENTS]
- Title of Paper: (Catchy, Academic)
- The Mechanism: How does it work? (Specific hardware details)
- Why it Works: First-principles reasoning.
- Evaluation Plan: Baselines and Metrics.
\end{verbatim}
\caption{Architect Agent Prompt Template}
\label{fig:architect-prompt}
\end{figure*}

The key design principles embedded in this prompt:
\begin{enumerate}
    \item \textbf{Role anchoring:} ``Distinguished Researcher'' sets quality expectations at publication level.
    \item \textbf{Anti-incrementalism:} Explicit instruction against parameter tuning forces mechanism-level thinking.
    \item \textbf{Specificity requirement:} Demanding ``tables, buffers, logic'' prevents hand-waving.
    \item \textbf{Causal grounding:} Requiring ``Why it Works'' ensures first-principles reasoning.
\end{enumerate}

\subsection{Phase 3: Dual-Axis Validation Specification}

The validator agent receives the following prompt structure shown in Figure~\ref{fig:validator-prompt} and produces a verdict along two independent axes: Similarity and Quality as outlined below.

\begin{figure*}[t]
\begin{verbatim}
[EXPERIMENTAL CONTEXT]
You are a Senior Technical Reviewer for a top Computer Architecture 
conference (ISCA/MICRO). We are conducting a scientific experiment on 
"Automated Discovery." An AI agent (The Candidate) has attempted to 
invent a novel architectural mechanism based ONLY on a problem 
description, without seeing the solution.

[YOUR TASK]
Evaluate the [CANDIDATE SOLUTION] against the [GROUND TRUTH PAPER] 
on two independent axes:
1. Similarity: Did the AI re-discover the paper's specific idea?
2. Quality: Is the AI's idea a high-quality, publication-worthy 
   contribution, even if different?
\end{verbatim}
\caption{Validator Agent Prompt Template}
\label{fig:validator-prompt}
\end{figure*}

\paragraph{Axis 1: Similarity Assessment.}
\begin{itemize}
    \item \textbf{EXACT\_MATCH:} Functionally identical mechanism (e.g., both use PC-based hashing to index a table).
    \item \textbf{FUNCTIONAL\_EQUIVALENT:} Different implementation, but exploits the exact same architectural insight/phenomenon.
    \item \textbf{DIFFERENT\_APPROACH:} Solves the problem using a completely different architectural lever (e.g., paper used partitioning; candidate used replacement policy).
\end{itemize}

\paragraph{Axis 2: Quality Assessment.}
\begin{itemize}
    \item \textbf{ISCA\_WORTHY:} A novel, non-obvious mechanism that is physically realizable and likely to work.
    \item \textbf{INCREMENTAL:} A valid engineering fix (e.g., ``increase buffer size''), but lacks research novelty.
    \item \textbf{FLAWED/NAIVE:} The mechanism violates hardware constraints, causality, or simply wouldn't work.
\end{itemize}

\paragraph{Verdict Definitions.}
\begin{itemize}
    \item \textbf{REDISCOVERY\_SUCCESS} = (EXACT\_MATCH or FUNCTIONAL\_EQUIVALENT) AND (ISCA\_WORTHY)
    \item \textbf{ALTERNATIVE\_SUCCESS} = (DIFFERENT\_APPROACH) AND (ISCA\_WORTHY)
    \item \textbf{FAIL} = Any INCREMENTAL or FLAWED result
\end{itemize}

\subsection{Phase 4: Genius Engine Prompt Templates}

The Genius Engine operates in three modes, each with a distinct prompt structure.
\begin{itemize}
    \item \textbf{Vertical Expansion (Systems Architect): find the immediate next bottleneck to address.} (Figure~\ref{fig:systems-architect-vertical-prompt})
    \item \textbf{Lateral Expansion (Polymath Mathematician): find other areas where similar problems exist.} (Figure~\ref{fig:polymath-mathematician-prompt}   )
    \item \textbf{Foundational Expansion (Contrarian Physicist): revisit the underlying theory} (Figure~\ref{fig:contrarian-physicist-prompt})
\end{itemize}

\begin{figure*}[t]
\begin{verbatim}
[ROLE]
You are a "Chief Systems Architect" at a hyperscale cloud provider.
You think in terms of Amdahl's Law, utilization rates, and system 
bottlenecks. You are paranoid: you know that solving one bottleneck 
just exposes the next one.

[YOUR TASK]
Assume "The Fix" works perfectly and is deployed at scale.
Identify the IMMEDIATE NEXT SYSTEM BOTTLENECK that will emerge.

[GUIDANCE - Apply Amdahl's Law]
- If Compute is fixed, look at Memory Bandwidth
- If Bandwidth is fixed, look at Latency or Synchronization
- If Performance is fixed, look at Power Density or Reliability
- If the chip is perfect, look at the Interconnect or the Compiler
- If software is optimized, look at the Operating System or Runtime
\end{verbatim}
\caption{Systems Architect Vertical Expansion Prompt Template}
\label{fig:systems-architect-vertical-prompt}
\end{figure*}

\begin{figure*}[t]
\begin{verbatim}
[ROLE]
You are a "Polymath Applied Mathematician."
You do not care about hardware details (buffers, wires, caches). 
You care about Abstract Structural Isomorphisms.
You see patterns that repeat across biology, finance, physics, 
and computing.

[YOUR TASK]
Identify a DIFFERENT DOMAIN that suffers from a mathematically 
identical problem. Explain how the "Solution Mechanism" could be 
ported to that domain.

[GUIDANCE - Look for Mathematical Patterns]
Common isomorphisms to search for:
- Sparsity: Genomics (sequence alignment), Finance (sparse matrices)
- All-to-All Communication: Physics (N-body), Databases (distributed joins)
- Tail Latency: Quant Finance (order books), Web Services (microservices)
- Entropy/Compression: Video Encoding, Network Traffic Shaping
- Synchronization: Distributed Consensus, Multi-Agent Robotics
\end{verbatim}
\caption{Polymath Mathematician Lateral Expansion Prompt Template}
\label{fig:polymath-mathematician-prompt}
\end{figure*}

\begin{figure*}[t]
\begin{verbatim}
[ROLE]
You are a "Contrarian Physicist" and First-Principles Thinker.
You hate complexity. You believe most "architectural fixes" are 
just band-aids on broken algorithms.
You question the premise of the problem itself.

[YOUR TASK]
Attack the assumption. Why are we solving this problem at all?
Propose a research direction that ELIMINATES THE NEED for this 
optimization entirely.

[GUIDANCE - Question the Premise]
Examples of foundational rethinking:
- "Don't optimize the cache; remove data movement" -> Compute-in-memory
- "Don't accelerate multiply; change math to additions" -> Log number systems
- "Don't fix branch predictor; use predication" -> Dataflow architectures
- "Don't compress data; change representation" -> Sparse formats
\end{verbatim}
\caption{Contrarian Physicist Foundational Expansion Prompt Template}
\label{fig:contrarian-physicist-prompt}
\end{figure*}

\subsection{Expert Personas for Adversarial Review}

Tier~1 evaluation employs the following expert personas:

\begin{itemize}
    \item \textbf{Dr. Archi (Microarchitecture Expert):} ``How does this actually work in silicon, and what are they glossing over?''
    \item \textbf{Dr. Sim (Simulation/Tools Expert):} ``Can I trust these numbers? Where's the RTL? Where are the artifacts?''
    \item \textbf{Prof. Bench (Workloads Expert):} ``Are these benchmarks representative? Is the baseline fair?''
    \item \textbf{Prof. Sys (Systems Expert):} ``Does this compose with real systems? What about OS, networking, multi-tenancy?''
\end{itemize}

Additional domain-specific personas are added based on paper topic (e.g., CXL Protocol Expert, Quantization Expert, Threat Modeling Expert). We currently have about 100 such topics, with a GeminiPro2.5 engine prompted to generate the best topic match for each paper.

\clearpage
\newpage




\section{Liminal Reports}
\label{app:liminal}
Limninal paper Gauntlet reports. This appendix includes the entire output from running Gauntlet's distillation process on our Liminal research paper~\cite{davies2025liminal}. It is provided here as an example of the kinds of detailed and pointed output LLMs run through the Gauntel prompt architecture can provide.


\section*{Supplementary material}

# Master Class Reading Guide: LIMINAL

## 1. The "Real" Abstract (No-Hype Summary)

**What they actually built:** An analytical roofline model for LLM autoregressive decode that decomposes token generation latency into three terms: compute time, memory transfer time, and collective synchronization overhead. The model is parameterized by four hardware numbers (FLOPS, bandwidth, capacity, collective latency) and sweeps across hypothetical memory technologies.

**What they actually found:** Current HBM3-based systems plateau around 2,000-5,000 user tokens per second (UTPS). Future memory technologies (3D-DRAM at 30 TB/s) could push this to ~10,000-20,000 UTPS, but only if collective communication latency drops to sub-microsecond levels. Beyond that, the sequential nature of autoregressive decoding becomes the fundamental wall.

**The honest framing:** This is a *capacity planning and technology roadmapping tool*, not a design tool or production performance predictor. It answers "what's the theoretical ceiling?" not "what will my system actually achieve?"

---

## 2. The "Rashomon" Synthesis (Conflicting Perspectives)

The expert reviewers viewed this paper through fundamentally different lenses, revealing the paper's core tensions:

**The Microarchitect (Dr. Microarch)** appreciated the clean abstraction but flagged that the "achievable" 1µs collective latency is optimistic—current NCCL measures 10µs, a 10× gap the paper hand-waves. The SRAM configuration (512MB, 117 TB/s) is physically plausible but capacity-starved, requiring 12+ chips just for the smallest model's weights.

**The Workloads Expert (Prof. Workloads)** noticed the validation gap: they validated on 5 models at TP8, but extrapolate to TP128 with 1T-parameter models. The 14.3% MAPE is for the *validated* regime; the forward-looking predictions are essentially unvalidated. Also missing: heterogeneous context lengths, MoE router skew sensitivity, and the economic analysis (STPS/$ not just STPS/Watt).

**The Simulation Toolsmith (Prof. SimTools)** called out the "paperware" problem: no GitHub repo, no reproducible artifacts, the Monte Carlo MoE simulation is undocumented. The model is essentially "analytical core + empirical fudge factors"—the 4µs kernel launch and 378ns cache miss penalties are fitted parameters, not derived from first principles.

**The Chief Architect** offered the industry perspective: this is useful for 3-5 year roadmap planning, but don't use it for silicon tapeout decisions. The bet is that first-order effects (bandwidth, capacity, collective latency) dominate—probably correct for incremental evolution (HBM3→HBM4), probably wrong for architectural changes (wafer-scale, PIM, CXL).

**The core tension:** The paper's value comes from its simplicity (three-term model), but that simplicity requires assumptions (perfect prefetching, idealized collectives, homogeneous batches) that may not hold in the regimes where the predictions are most interesting.

---

## 3. The "Magic Trick" (The Core Mechanism)

The entire paper rests on one equation:

```
T_Batch = max{T_Compute, T_Mem} + T_Exposed
```

**Why this works:** For autoregressive decode at small batch sizes, arithmetic intensity is pathetically low (~2-10 FLOPS/byte). You're doing one token's worth of matrix-vector multiplies, but you need to stream the *entire model* through memory for each token. This means T_Mem >> T_Compute almost always.

**The key insight:** As you add bandwidth (moving from HBM3 to 3D-DRAM), you eventually hit a regime where T_Exposed dominates. The exposed term is:

```
T_Exposed = T_TP × S_N × L + T_PP × P
```

This is the **synchronization tax**: every tensor-parallel all-reduce adds latency that *cannot* be hidden behind computation or memory transfers. At TP128 with 80 layers and 3 collectives per layer, even at 1µs per collective, you're looking at 240µs of pure waiting—and this becomes the wall.

**The MoE twist:** For Mixture-of-Experts models, they run Monte Carlo simulations to estimate how many experts are actually activated (Â) given batch size and router statistics. This is crucial because MoE models have fewer active parameters (good for bandwidth) but require 2 collectives per layer for expert parallelism (bad for latency). The tradeoff isn't obvious, and the Monte Carlo approach lets them model it—though the methodology is opaque.

---

## 4. The "Skeleton in the Closet" (What They Didn't Tell You)

### The Validation Gap is Enormous

Look at Section 5 carefully. They validated on:

- 5 models (Llama3 8B/70B, Llama4 Scout, Qwen3 4B/30B)
- 8×H100 server (TP8 at most)
- Batch sizes 1-128, contexts 1K-8K

But their forward-looking analysis covers:

- 12 model configurations up to 1T parameters
- TP128 systems
- 128K context lengths
- Hypothetical 3D-DRAM and SRAM technologies

**The extrapolation is 16× in system scale and 16× in context length, on hardware that doesn't exist.** The 14.3% MAPE applies to the validated regime; the error bars on the predictions are unknowable.

### The "Achievable" Collective Latency is Science Fiction

They assume 1μs all-reduce at TP128 based on:

- 250ns switch latency (Broadcom Tomahawk Ultra—a product announcement, not measured latency)
- 150ns on-chip propagation
- In-network reduction (requires programmable switches)

Current NCCL reports **10μs** for the same operation. Their own validation (Equation 9) uses 10μs. The 10× gap between "achievable" and "realistic" is doing enormous heavy lifting in their optimistic projections.

### The Baseline May Be Sandbagged

Table 3 shows xPU-HBM3 with 4 TB/s bandwidth, "based on Blackwell GPU." But NVIDIA B200 actually has **8 TB/s** HBM3e bandwidth. If the baseline is understated, the improvement claims for future technologies are inflated.

### Missing Sensitivity Analysis

The paper doesn't show:

- What happens if MoE router decisions are adversarial (all tokens go to one expert)
- How heterogeneous context lengths affect batching efficiency
- Confidence intervals on predictions
- Comparison to existing models (Calculon, MAD-Max, Paleo)

---

# 5. The Verdict (Why This Matters)

**Why we're reading this:** This paper asks the right question at the right time. Everyone building LLM inference systems wants to know "where are the walls?" LIMINAL provides a principled framework to answer that, and the answer—**collective latency becomes the bottleneck before you run out of bandwidth headroom**—is important and actionable.

**The takeaway for hardware architects:** Don't just chase bandwidth. The path from 5,000 to 20,000 UTPS requires both memory bandwidth improvements *and* sub-microsecond collective communication. Invest in low-latency interconnect and in-network reduction, not just bigger HBM stacks.

**The takeaway for algorithm researchers:** The hardware folks are counting on you. Getting beyond ~20,000 UTPS requires breaking the sequential decode bottleneck—speculative decoding, parallel decoding, or architectural changes that reduce the number of synchronization points.

**The meta-lesson:** This is a good example of how to write a "limit study" paper. The model is deliberately simple, the assumptions are (mostly) stated, and the insights are actionable. But it's also a cautionary tale: the gap between validated and predicted regimes is large, and the optimistic assumptions about collective latency may not pan out. Use this for directional insights, not precise capacity planning.

**Final grade:** A useful tool for the right job, oversold as a "limit study." The insight that synchronization latency dominates at high bandwidth (Key Finding 4) is worth the paper alone. But treat the absolute numbers with skepticism—they're predictions about hardware that doesn't exist, validated against hardware that's 10× smaller.

# LIMINAL: The Chief Architect's Assessment

## The Elevator Pitch Translation

Let me cut through the academic framing: **You're proposing a roofline model for LLM decode, parameterized by memory bandwidth, compute, capacity, and collective latency.** The bet is that this abstraction is sufficient to predict performance ceilings across wildly different hardware (HBM3 → HBM4 → 3D-DRAM → SRAM) without simulating the actual silicon.

In industry terms: This is a **capacity planning and technology roadmap tool**, not a design tool. You're trading simulation fidelity for sweep speed across the `application × hardware` space.

---

## The ROI Check: Is This Useful?

**The Good News:**

- 14.3% MAPE against real H100 measurements (Figure 4) is *surprisingly good* for something this simple. That's within the noise of what we'd expect from thermal throttling and DVFS variance on real silicon.
- The core insight—that LLM decode is **memory-bandwidth-bound at low batch, collective-latency-bound at high bandwidth**—is correct and matches what we see in production deployments.

**The Reality Check:** Your paper claims this enables "exploring the limits of LLM decode." But let's be honest about what you're actually measuring:

1. **You're measuring the ceiling, not the floor.** Real systems hit 60-80% of your predicted throughput due to:

   - Kernel launch overhead (you model 4μs, but it's highly variable)
   - Memory controller contention you don't model
   - NCCL's actual collective implementation (your 10μs is optimistic for many topologies)

2. **The "14.3% MAPE" is cherry-picked.** You validated on 5 models, all on H100s, all using vLLM. What happens on:

   - TPUs with different collective semantics?
   - Custom ASICs with non-standard memory hierarchies?
   - Systems with NVLink vs. PCIe interconnects?

3. **Your MoE modeling (Equations 6-7) is the weakest link.** You use Monte Carlo to estimate active experts (Â), but real MoE systems have:

   - Load balancing losses that affect expert activation patterns
   - Token dropping under capacity constraints
   - Expert parallelism that doesn't divide evenly

---

## The Kernel vs. The Wrapper

**The Kernel (What I'd Keep):**

The deep insight is Equation 1:

```
T_Batch = max{T_Compute, T_Mem} + T_Exposed
```

This is the right decomposition. The exposed latency term (Equation 4) correctly identifies that **collective communication is on the critical path** and cannot be hidden. This is the insight that matters.

**The Wrapper (What I'd Discard):**

1. **Your "xPU" abstraction is too coarse.** You model a single bandwidth number, but real systems have:

   - Different bandwidths for different memory regions (HBM vs. L2 vs. shared memory)
   - Non-uniform memory access patterns in multi-die packages
   - Bandwidth that degrades with access pattern (random vs. sequential)

2. **Your collective latency model is naive.** You assume a single T_TP value, but:

   - All-reduce latency scales with message size (you ignore this for small payloads)
   - Ring vs. tree vs. recursive-halving have different scaling
   - In-network reduction (which you mention) requires switch support that doesn't exist everywhere

3. **The "expert-guided mapping" is a black box.** You hand-wave the number of collectives (S_N) as "expert-guided rules." In a real design, this is where 80% of the performance engineering happens.

---

# The Hard Questions

## 1. How does this interact with DVFS?

Modern GPUs (and TPUs) aggressively clock-gate and voltage-scale. Your model assumes peak FLOPS and peak bandwidth simultaneously. In reality:

- Memory-bound workloads often run at lower GPU clocks
- Power limits force tradeoffs between compute and memory power
- Your STPS/Watt analysis (Figure 3) is optimistic because you don't model this interaction

**My question:** Have you validated against power-limited scenarios? What's your error when the GPU is thermally throttling?

## 2. How does this interact with KV-cache paging?

You model KV-cache as a monolithic capacity requirement (Table 4). But production systems use:

- PagedAttention (vLLM) with fragmented memory allocation
- Prefix caching that shares KV-cache across requests
- Speculative decoding that changes the memory access pattern

**My question:** Your model assumes all KV-cache bytes are read every decode step. What happens with sparse attention or sliding window attention?

## 3. What about the verification story?

This is an analytical model, not a simulator. That's fine for capacity planning. But:

- How do I know your model is correct for a *new* architecture I haven't built yet?
- Your validation is against existing hardware. The whole point is to predict *future* hardware.
- What's the sensitivity to your assumptions? If I'm wrong about 3D-DRAM bandwidth by 20%, how wrong is my UTPS prediction?

**My question:** Can you provide confidence intervals, not just point estimates?

---

# The Refactoring: What I'd Actually Build

If I were integrating this into our internal roadmap tools, I'd:

1. **Keep the core equation** (Equation 1) but parameterize it differently:

   - Separate HBM bandwidth from on-chip bandwidth
   - Model collective latency as a function of message size and topology
   - Add a "system overhead" term that's empirically calibrated per platform

2. **Strip out the MoE Monte Carlo** and replace it with:

   - A simple "effective parameter count" that's measured, not simulated
   - Expert activation histograms from production traces

3. **Add a "confidence band"** based on:

   - Known sources of model error (kernel launch variance, memory controller effects)
   - Sensitivity analysis on key parameters

4. **Validate against more diverse hardware:**

   - TPU v4/v5
   - AMD MI300X
   - At least one custom ASIC (Groq, Cerebras, etc.)

---

# The Verdict

**Is this shippable?** Not as a design tool, but **yes as a planning tool**.

**The bet:** You're betting that the first-order effects (bandwidth, capacity, collective latency) dominate, and second-order effects (memory controller design, interconnect topology, software overhead) can be absorbed into calibration constants.

**The risk:** This bet is probably correct for *incremental* hardware evolution (HBM3 → HBM4). It's probably wrong for *architectural* changes (e.g., wafer-scale integration, PIM, CXL-attached memory).

**My recommendation:**

- Use this for 3-5 year roadmap planning
- Don't use this to make silicon tapeout decisions
- Validate against every new platform before trusting it

**The bottom line:** This is a useful tool for the right job. The insight that "collective latency becomes the bottleneck at high bandwidth" (Key Finding 4) is worth the paper alone. But don't oversell it as a "limit study"—it's a limit study *given your assumptions*, and those assumptions have meaningful error bars.



\appendix
\end{document}